\documentclass[epj]{svjour}
\usepackage{graphicx}

\usepackage{amsmath,amssymb,mathdots}

\usepackage[enableskew]{youngtab} %enableskew  stdtext 

\usepackage{graphics}
\usepackage{float}
\usepackage{longtable}
\usepackage{supertabular}
\usepackage{color}
\usepackage{lipsum}

\def\be{\begin{equation}}
\def\ee{\end{equation}}

\begin{document}
\title{An alternative approach to the construction of Schur-Weyl transform}
\author{Pawe\l\enskip Jakubczyk\inst{1} \and Yevgen Kravets\inst{2} \and Dorota Jakubczyk\inst{3}%
%\thanks{\%emph{Present address:} Insert the address here if needed}%
}                     % Do not remove
%
%\offprint%s{}          % Insert a name or remove this line
%
\institute{Faculty of Mathematics and Natural Sciences, University of Rzesz\'{o}w, Rejtana 16A, 35-959 Rzesz\'ow, Poland
\and Centre de Physique Th\'{e}orique (CPHT), \'{E}cole Polytechnique, 91128 Palaiseau, France 
\and Department of Physics, Rzesz\'{o}w University of Technology, al. Powsta\'{n}c\'{o}w Warszawy 12, 35-959 Rzesz\'{o}w, Poland}
\date{Received: date / Revised version: date}
\abstract{
We propose an alternative approach for the construction of the unitary matrix which performs generalized unitary rotations of the system consisting of independent identical subsystems (for example spin system). This matrix, when applied to the system, results in a change of degrees of freedom, uncovering the information hidden in non-local degrees of freedom. 
This information can be used, inter alia, to study the structure of entangled states, their classification and may be useful for construction of quantum algorithms.
} %end of abstract
\maketitle
\section{Introduction}

There is a variety of approaches to the problem of classification of the entangled states.  
Based on the fact that not every entangled state has the same properties an attempt has been made to classify them using local hidden variables \cite{horodecki1,nr2,nr3,nr4,nr5,nr6}, or using SLOCC (Stochastic Local quantum Operations and Classical Communications) invariance of the states by distributing them among different classes of equivalence \cite{nr7,nr8,nr9,nr10}. The newest approach (although indirectly related to the SLOCC invariance) is based on the classification by assigning the entangled state to a defined multidimensional geometrical object (so called \textit{polytope}) \cite{nr11}.

However, another property of quantum entanglement, which is not widely considered, can prove to be useful to the procedure of states classification: entanglement is not preserved with the change of degrees of freedom describing the system \cite{nr12}. 
Change in the degrees of freedom of the system can reveal information stored in non-local degrees of freedom and expose partially (or fully) internal structure of quantum entanglement within this system. 
Required changes to the degrees of freedom can be performed in terms of generalized unitary rotations of the system by acting on it with a unitary matrix.
This matrix is identified in the literature as the Schur-Weyl (SW) transform \cite{nr14} due to being based on the Schur-Weyl duality \cite{schur,weyl}.
In \cite{nr13} we have shown that application of one magnon SW transform  to small spin system (couple qubits) reveals the structure of bipartite entanglement. This information provided classification for all types of bipartite entanglement in terms of standard Young tableaux. Moreover, in \cite{nr13a} we have shown that by coupling SW transform with the KKR algorithm \cite{KKR_comp} classification of entanglement states can then be done via rigged string configuration~\cite{KKR}.

In this paper we propose an alternative method for constructing the SW transform based on fundamental tensor operators \cite{louck,louck4,louck4a}. This method exploits the symmetry of the model and can be applied to systems consisting of independent constituents. 
This work is restricted to the one dimensional spin systems, however our approach can be extended to every system consisting of identical subsystems.

The remainder of this paper is organised as follows. Section 2 starts with a brief description of the model: we introduce the irreducible basis states of the SW duality and establish the connection between them. In Section 3 we present the method of construction of matrix elements of the SW transform. Section 4 contains an example and we conclude in Section 5 with a brief summary of our results.

\section{The model}

Consider one-dimensional spin chain consisting of $N$ nodes, each with spin $s$. 
Basis elements of the Hilbert space for this model are given by the set of single-node product states of the form
\be\label{productBasis} 
\begin{array}{l}
|f\rangle = | i_1 \rangle \otimes | i_2 \rangle \otimes \ldots \otimes | i_N \rangle \equiv | f(1), f(2), ..., f(N) \rangle,\\
i_j \in \tilde n,\;  j \in \tilde N,\\
\end{array}
\ee
where $\tilde{n}=\{i = 1,2,...,n\}, \,\,\, n=2s+1$ denotes single nodes states and $\tilde{N}=\{j = 1,2,...,N\}$ the set of nodes of the model.
Set $\tilde n^{\tilde N}$ of all product states forms the computational basis which spans unitarly the Hilbert space
$
\mathcal{H} = lc_{\mathbb{C}} \,\, \tilde n^{\tilde N}
$
of the system. 

By definition, this system reveals the symmetry under collective unitary \emph{rotations} $u \in U(n)$ in single-node spaces, and well defined transformation properties under the permutations of nodes $\sigma \in \Sigma_N$.

Basically the space $\mathcal H$ is a scene of two dual actions:
$A: \Sigma_N \times \mathcal{H} \rightarrow \mathcal{H}$ and
$ B: U(n) \times \mathcal{H} \rightarrow \mathcal{H}$, which are the symmetric and unitary group actions, respectively.
These two actions mutually commute i.e.
$[A(\sigma),B(u)]=0, $ for $\sigma \in \Sigma_N,$ and  $u \in U(n),$
despite the fact that both groups for $N>2, n>1$ are highly non-commutative. This commutation relation is a source of SW duality. 

To describe SW duality we introduce the notion of \emph{partition} \cite{fulton,sagan}. The partition $\lambda$ of the number $N$ into $n$ parts is the sequence of numbers (parts) $( \lambda_1, \lambda_2, \ldots, \lambda_n )$ which fulfil the following conditions
\begin{equation}
\lambda_1 \geq \lambda_2 \geq \ldots \geq \lambda_n \geq 0\ ,
\end{equation}
and 
\begin{equation}
\sum_{i \in \tilde n} \lambda_i = N.
\end{equation}
The symbol $D_W(N,n) $ denotes the set of all partitions of the number $N$ into no more than $n$ parts. 
Partitions $\lambda$ serve as labels for classifications of irreducible representations (irreps) $D^\lambda$ and $\Delta^\lambda$ of the unitary $U(n)$ and symmetric $\Sigma_N$ groups, respectively.

In terms of partition $\lambda$ one can write decomposition of actions $A$ and $B$ into irreps
\be\label{sw1}
A=\sum_{\lambda \in \mathcal D_W(N,n)}\,\, m(A,\Delta ^{\lambda})\,\,\Delta ^{\lambda},
\ee
\be\label{sw2}
B=\sum_{\lambda \in \mathcal D_W(N,n)}\,\, m(B,D^{\lambda})\,\,D^{\lambda},
\ee
of the symmetric and unitary groups.
Here $m(A,\Delta ^{\lambda})$ denotes multiplicity of occurrence of the irrep $\Delta ^{\lambda}$ in representation $A$, while $m(B,D^{\lambda})$ multiplicity of occurrence of irrep $D^{\lambda}$ in representation $B$.
The appropriate multiplicities on the strength of SW duality satisfy the following relations
\be\label{krotnosci}
m(A,\Delta^\lambda)=dim \, D^{\lambda}, \quad\quad
m(B,D^\lambda)=dim \, \Delta^{\lambda},
\ee
where $\lambda \in \mathcal D_W(N,n)$, and the symbol $dim$ stands for dimension of the representation.
This way the SW duality decomposes the entire space $\mathcal H$ of quantum states of the composite system into \emph{sectors} $\mathcal{H} ^{\lambda}$ 
\begin{equation}\label{r19}
\mathcal{H} = \sum_{\lambda \in  \mathcal D_W(N,n)} \oplus\,\, \mathcal{H} ^{\lambda}
\end{equation}
labelled by partitions $\lambda$. 

In order to find the irreducible bases in the sectors $\mathcal{H} ^{\lambda}$ let us first introduce the irreducible bases labels: the standard Young tableau and the semistandard Weyl tableau  \cite{fulton,sagan}.

Standard Young tableau $y$ can be regarded as a bijective mapping $y : sh \, \lambda \longrightarrow \tilde N$ of the set of boxes of the diagram of $\lambda$ to the alphabet $\tilde N$  of nodes,  which satisfies \emph{standardness conditions}
\begin{equation}
 \alpha' > \alpha \Rightarrow y_{\alpha' \beta} > y_{\alpha \beta}, \,\,\,\,\,\,\,\,\,\,\,\, \beta' > \beta \Rightarrow y_{\alpha \beta'} > y_{\alpha \beta},
\end{equation}
where $\alpha$ denotes row number and $\beta$ column number of the element $y_{\alpha \beta}$ in the Young diagram $sh \, \lambda$\footnote{Young diagram $sh \, \lambda$ (\textit{sh} is abbreviation of shape) denotes empty Young or Weyl tableau i.e. shape of tableau.}.
This effectively means that entries $y_{\alpha \beta} \in  \tilde N$ strictly increase along each row and each column of the Young diagram. The set of all standard Young tableaux of the shape $\lambda$ on the alphabet of nodes is given by $SYT(\lambda, \tilde N)$. 

A semistandard Weyl tableau $t$ on the other hand is a mapping $t: sh \, \lambda \rightarrow \tilde n$ with the same domain $sh \, \lambda$ and the alphabet $\tilde n$ of spins as the target, with \emph{semistandardness conditions}
\begin{equation}
 \alpha' > \alpha \Rightarrow t_{\alpha' \beta} \geq t_{\alpha \beta}, \,\,\,\,\,\,\,\,\,\,\,\, \beta' > \beta \Rightarrow t_{\alpha \beta'} > t_{\alpha \beta},
\end{equation}
so that entries $t_{\alpha \beta} \in \tilde n$ do not decrease (or weakly increase, with possible repetitions) along each row, and strictly increase along each column of $sh \, \lambda$.
The set of all semi-standard Weyl tableaux of the shape $\lambda$ on the alphabet of spins is denoted by $SSWT(\lambda, \tilde n)$.

According to (\ref{sw1},\ref{sw2}) and (\ref{krotnosci}), each sector $\mathcal H^\lambda$ can be factorized into irreducible sectors 
\begin{equation}\label{r22}
\mathcal H^\lambda = U^\lambda \otimes V^\lambda,
\end{equation}
where  $U^\lambda = lc_{\mathbb{C}} SSWT(\lambda,n)$ is the carrier space of the irrep $D^{\lambda}$ and $V^\lambda = lc_{\mathbb{C}} SYT(\lambda)$ is the carrier space of irrep $\Delta^{\lambda}$.
Decomposition (\ref{r22}) justifies introducing the irreducible basis labels $t \in SSWT(\lambda,n)$ for the representation $B(u)$ and $y \in SYT(\lambda)$ for $A(\sigma)$ in accordance with representation theory of symmetric and unitary groups. 

This approach defines a new irreducible basis for the model, which reads
\begin{align}\label{sws}
\nonumber b_{irr} = &\{ | \lambda \, t \, y \rangle \; : \; \lambda \in  \mathcal D_W(N,n), \, \\
&\quad \quad t \in SSWT(\lambda, \tilde n), \, y \in SYT(\lambda, \tilde N) \}.
\end{align}
Each sector $\mathcal H^\lambda$ disposes a separable basis $| t \, y \rangle$, where $t$ and $y$ are associated with local and global variables respectively.
Introduction of the basis (\ref{sws}) resulted in the new choice of degrees of freedom. For the product basis (\ref{productBasis}) we initially had $N$ degrees of freedom, whereas for the SW basis (\ref{sws}) the number of degrees of freedom was reduced to three. 
Due to the orthogonality properties of these bases we expect the following unitary transformation between them
\begin{equation}\label{rozKostki}
|\lambda \, t \, y \rangle \hspace{-3pt} = \hspace{-5pt} \sum_{f \in {\tilde n^{\tilde N}}}
\langle f | \lambda t y \rangle \;
|f \rangle,
\end{equation}
with coefficients $\langle f | \lambda t y \rangle$ forming a unitary matrix. This matrix transforms the initial basis $\tilde n^{\tilde N}$ of  product states with fixed decomposition of spins projections into the irreducible one of the SW duality. We refer to Eq. (\ref{rozKostki}) as the SW transform, which converts the initial base of product states $\tilde n^{\tilde N}$ into the irreducible base
$
\{ |\lambda \, t \, y \rangle \}
$
of the SW duality. Here rows of this matrix are indexed with $f$ and columns are identified using $(\lambda \, t \, y )$ triads.

\section{Construction of matrix elements of the SW transform}

According to general rules of quantum mechanics any operator $F$ which undergoes unitary transformation follows
$
F \rightarrow F' \equiv u F u^{-1}.
$

Term \emph{tensor operator} usually corresponds to a set of operators $\{F_t\}$, which transform irreducibly under any unitary transformation  i.e. transforming linearly into themselves
\be\label{jot1}
u F_t^\lambda u^{-1} = \sum_{t'} D_{t't}^\lambda(u) F_{t'}
\ee
where $D_{t't}^\lambda(u)$ is a matrix element  $t't$ of irreducible representation $D^\lambda$ for the element $u \in U(n)$.
It follows that operators transform under the same rules as basis elements of the irreducible representations of $U(n)$ group, and therefore span the carrier space of these representations. 

On the other hand the idea of tensor operators is closely associated with the Littlewood - Richardson decomposition $D^\lambda \otimes D^\mu = \sum_\nu c_{\lambda \mu}^\nu D^\nu$, with which the problem of multiplicity of the classification of states was solved. Here irreducible tensor operators create a carrier space of representation $D^\lambda$, whereas irreducible states of the system span the representation space of $D^\mu$. Therefore the product of an irreducible tensor operator $D_t^\lambda$ acting on the state $|\mu\, m \rangle$ is given by the Wigner-Eckart theorem for the matrix elemets
\be\label{jot1b}
\langle \nu\, n |D_t^\lambda| \mu\, m \rangle =
\sum_{\gamma=1}^{c_{\lambda \mu}^\nu} \big\langle \nu || D^\lambda|| \mu \big \rangle_\gamma
\left<
\begin{array}{@{}cccc@{}}
  \lambda & \mu & \nu & \gamma \\
  t & m & n &  \\
\end{array}\right \rangle
\ee
where $\big\langle \nu || D^\lambda|| \mu \big \rangle_\gamma$ corresponds to a reduced matrix element,
$
{\scriptsize
\left<
\begin{array}{@{}cccc@{}}
  \lambda & \mu & \nu & \gamma \\
  t & m & n &  \\
\end{array}\right \rangle}
$ is a Wigner-Clebsch-Gordan (WCG) factor for unitary groups and the sum runs over all the repetition indexes $\gamma$ of the irreducible representation $D^\nu$ in the product $D^\lambda \otimes D^\mu$.

Fundamental tensor operators \cite{louck,louck4,louck4a} are the simplest and the most important tensor operators as they allow us to build almost every other objects in the unitary group theory from them. They transform according to the fundamental representation $D^{(1)}(u)$ of $U(n)$.
For the group $U(n)$ there are $n$ fundamental tensor operators $t_{\bullet,\tau}$, $\tau \in \tilde n$ each with $n$ componets. 
Therefore one can introduce the notation $t_{k,\tau}$, where $k,\tau \in \tilde n$. 
The operator $t_{k,\tau}$, can be used as a toll which mathematically describes enlarging of our system by adding one node. More precisely, it describes the addition of the node with the number $k$ according to the symmetry marked by $\tau$. 

Taking into account the above observation the elements of the matrix (\ref{rozKostki}) can be obtained via expansion of the system by progressively adding the nodes as a one-by-one process consistent with the symmetry governed by the set of partitions
 $\{\lambda_{1}, \lambda_{12}, \ldots, \lambda_{1..N}=\lambda \}$, given by the RSK algorithm \cite{robinson,schensted,knuth}. Namely, 
$\lambda_{1}$ corresponds to the shape of Young tableau after inserting the letter (spin of node 1) $i_1$ into empty tableau, 
$\lambda_{12}$ corresponds to the shape of Young tableau after inserting the second letter (spin of node 2) $i_2$ into previous Young tableau, etc. 
Here, partition $\lambda_i$ describes the symmetry of the system consisting of $i$ nodes  and $\lambda$ decribes symmetry of the total system.

Addition of a single node, with the number $i$ to an existing system consisting of $i-1$ nodes in a state given by $|\lambda_{1\ldots j-1}\, t_{1\ldots j-1} \rangle$, leading to the final state $|\lambda_{1\ldots j}\, t_{1..j} \rangle$, is described by the fundamental tensor operator 
\begin{equation}\label{fop}
\langle \lambda_{1\ldots j}\, t_{1..j} |  \hat F_{f(j), row(\lambda_{1..j} \setminus \lambda_{1..j-1})}  |\lambda_{1\ldots j-1}\,  t_{1..j-1} \rangle
\end{equation}
where first subscript $f(j)$ of $\hat F$  
corresponds to the state of the added node $j$ and the second subscript $row(\lambda_{1..j} \setminus \lambda_{1..j-1})$ represents the row number of the partition $\lambda_{1..j}$ which hosts the new cell after adding the node $j$; here $|\lambda_{1\ldots j-1}\, t_{1..j-1}\rangle$ denotes the state of the system prior to the addition of the node $j$, and $|\lambda_{1\ldots j}\, t_{1..j}\rangle$ represents the final state.

As mentioned earlier construction of the coefficient $\langle f | \lambda t y \rangle$ in (\ref{rozKostki}) is based on the process of adding successive nodes to the system and can be described by fundamental tensor operators starting from the one-node system. This process corresponds to combinatorial growth of the shape of the Weyl tableau $t$, referred to hereafter as the process of \emph{crystallisation of the state} $| \lambda t y \rangle$ from the initial configuration $f$. Each such growth obeys selection rules given by the set of partitions $\lambda_{1}, \lambda_{12}, \ldots, \lambda_{1..N}$ at every stage, and thus contributes additionally to the total value of the coefficient, consistent with the quantum mechanical prescriptions of interference. 
The source of quantum interference is the fact that addition of the new node to the system can be performed in a variety of different ways which cannot be ''observed''. 
This results in "quantum interference" of different ways of adding the node.
Clearly, the entire building process can be represented by a path on the related graph, and the final result is the sum over all such paths.

To obtain systematic description of all possible ways of growth of the $t$ tableau representing the state of the system, graph $\Gamma$ for a given matrix element $\langle f| \mu \lambda t y \rangle$ has to be constructed by adding the successive letters of configuration $f$ to the Weyl tableau adjusted to the sequence $\lambda_{1}, \lambda_{12}, \ldots, \lambda_{1..N}$.

To simplify our procedure we exploit bijection between Weyl tableaux and Gelfand-Tsetlin patterns \cite{gelfand,gelfand1,louck,louck4} since both contain the same amount of information about the state of the system.
Our personal preference for this case lies with Gelfand-Tsetlin patterns due to the fact that their geometric construction (the betweenness conditions) provides better description of the growth of the system in the process of adding new nodes.

Gelfand-Tsetlin pattern is a triangular tableau $n(n+1)/2$ of non-negative integers $\{ m_{i, j} \}$, satisfying the betweenness conditions $m_{i-1, j} \leq  m_{i-1, j-1} \leq m_{i j}$ for all $ 1 \leq i \leq j \leq n$
\be\label{gu5}
\begin{array}{@{}lllllllll@{}}
m_{1n} &          & m_{2n}   &           & \cdots  & m_{n-1 n} &             & m_{nn}  \\

       & m_{1n-1} &          & m_{2 n-1} &         & \cdots    & m_{n-1 n-1}      &     \\

       &          & \ddots   &           &  \vdots &           & \iddots          &      \\

       &          & m_{1 3}  &           & m_{2 3} &           & m_{33}          &      \\

       &          &          & m_{1 2}   &         & m_{2 2}                       & &   \\

       &          &          &           & m_{1 1}                                 && &   \\
\end{array}.
\ee
These triangles classify all the basis states of the irreducible representations of the unitary group $U(n)$  labelled by partition from the $n$-th row $m_{1n}, m_{2n} \ldots m_{n-1 n}, m_{nn}$ of the triangle.

To be able to uniquely determine an irreducible representation to which the state belongs we will separate out the $n$-th row of the triangle 
\be\label{gu7}
\left (
\begin{array}{@{}c@{}}
  [m]_n \\
  (m)_{n-1} \\
 \end{array}
\right )
 \ee
in such a way that $[m]_n$ is a partition corresponding to an irreducible representation in the group $U(n)$ and $(m)_{n-1}$ corresponds to the remaining $n-1$ rows of the triangle.
Further we will identify the Weyl tableaux with Gelfand triangles and use them alternatively, taking into account the mentioned bijection.

Formally, graph $\Gamma$ consists of the set $GT$ of Gelfand-Tsetlin patterns
as vertices and the set $\{f(i):~~i =1 \ldots N\}$ of single-node states which label the edges (or arcs), such that $\Gamma = (GT, \{f(1), f(2), \ldots, f(N)\})$.
Edge $f(j)$  of two adjacent vertices $(t_{12..j-1}, t_{12..j})$, with $t_{12..j-1}$ being the initial and $t_{12..j}$ the terminal vertex, is constructed by inserting the single node state (the letter) $f(j)$  into the initial vertex $t_{12..j-1}$ to obtain the state $t_{12..j}$.
Such a graph is simple and directed with minimal (initial) vertex equal to the zero Gelfand-Tsetlin pattern, and maximal (final) vertex equal to the pattern corresponding to the Weyl tableau $t$.

More precisely, the process of construction of the graph $\Gamma$ can be split into two stages:

{\bf Stage 1}.--- We read off the sequence of partitions $\lambda_{RS} = (\lambda = \lambda_{12\ldots N}=[m]_n, \lambda_{12\ldots N-1}=[m]_{N-1}, \ldots, \lambda_{12}=[m]_2, \lambda_{1}=[m]_1)$ which corresponds to the growth of the shape of the Weyl tableaux in terms of the Robinson-Schensted-Knuth algorithm \cite{robinson,schensted,knuth} applied to the configuration $f$.
These partitions create the rows of the maximal Gelfand-Tsetlin pattern i.e. $[m]_j$ is the $j$-th row of the maximal Gelfand-Tsetlin pattern.

{\bf Stage 2}.--- The principal construction of the graph is based on an \emph{insertion} of the consecutive letters $f(j), \; j=1,2, \ldots, N$ of configuration $f=|f(1) f(2) \ldots f(N) \rangle$ to the Gelfand-Tsetlin patterns, starting with a triangle consisting of zeros only.
Insertion of a letter $f(j)$ into Gelfand-Tsetlin pattern $t_{1..j-1}$ leads to the pattern $t_{1..j}$ with new elements increasing by one and being located in the rows $j$,~~$f(j)~\leq~j~\leq~n$, i.e.
\begin{equation}\label{k8}
\left (
\begin{array}{c}
  [m]_n + e_n(\tau_n) \\
  \mbox{[$m$]}_{n-1} + e_{n-1}(\tau_{n-1}) \\
  \vdots \\
   \mbox{[$m$]}_{f(j)} + e_{f(j)}(\tau_{f(j)})\\
  (m)_{{f(j)}-1}\\
   \end{array}
 \right )
\end{equation}
where $\tau_j \in \{ 1,2, \ldots j \}$
for $j=f(j),f(j)+1, \ldots n-1, n$,
and $[m]_j + e_j(\tau_j)$ correspond to row $j$ of the Gelfand-Tsetlin pattern, $e_j(\tau_j)$ denotes zero vector of the length $j$ with $1$ at the position $\tau_j$.
The symbol $(m)_{{f(j)}-1}$ represents rows of Gelfand pattern numbered from $1$ to ${f(j)}-1$. 

It is obvious that this operation can lead to a collection of patterns (due to $\tau_j \in \{ 1,2, \ldots j \}$), but we choose only those, for which the $n$-th row $[m]_n + e_n(\tau_n)$ is equal to a partition $\lambda_{12..j}\in \lambda_{RS}$, and the standardness of the Gelfand-Tsetlin pattern is conserved (the betweenness conditions are satisfied).
In terms of graphs this corresponds to the possibility for an out-degree of the vertex (i.e. the number of edges coming from vertex) $t_{1..j-1}$ denoted by $deg^+(t_{1..j-1})$ to be greater than (or equal to) one.

In summary this can be implemented as follows

{\bf Step 1}.--- Using the above insertion procedure we start to insert the first letter $f(1)$ into the zero Gelfand triangle $t_0$ (i. e. a triangle of the shape $\lambda$, filled in by zeroes, which is the minimal vertex of our graph), which results in reaching the vertex $t_1$. This leads to a directed graph, consisting of two vertices $(t_0, t_1)$, joined by the edge $f(1)$
$$
\begin{array}{ccc}
&(t_0)&\\
&\downarrow &{\tiny f(1)}.\\
&(t_1)&\\
\end{array}
$$

{\bf Step 2}.--- Next, one inserts the letter $f(2)$ into the triangle $t_1$ leading to a set of vertices $t_{12}=\{t_{12}^i \; : \; i=1,2,...\}$. Geometrically this represents a graph exhibiting branches, with $deg^+(t_1) \geq 1$.
$$
\begin{array}{ccccc}
&&(t_0)&&\\
&&\;\;\;\;\;\;\;\; \downarrow {\tiny f(1)}&&\\
&&(t_1)&&\\
\;\;\;\;\;\;\;\;\;\;\;\; \swarrow f(2)&\ldots &\;\;\;\;\;\;\;\;\downarrow  f(2)& \ldots &\\
(t_{12}^1)&\ldots &(t_{12}^k)& \ldots & \ldots \\
\end{array}
$$

{\bf Step 3}.--- This is followed by further insertion of the letter $f(3)$ into each vertex $t_{12}^i$ from the set $t_{12}$ using the same rules, which produces the set $t_{123}$ of vertices composed of three letters. The same routine is followed for all remaining letters of the configuration $f$. 

One can observe that the out-degree $deg^+(t_{12..j})\geq 1$, and can be seen that the insertion rules themselves suggest a quick growth of the graph in a tree-like manner. Nevertheless, symmetry constraints imposed by the physical system, guarantee that final graph will result in the shape of a rhomb (see example below) with the maximal (final) vertex resulting from the insertion of the last letter $f(N)$ of the configuration $f$ equal to $\lambda t$.

The amplitude of such graph $\Gamma$ corresponds to (\ref{wsp})
\begin{figure*}
  \hrulefill
  \begin{equation}\label{wsp}
\langle f | \lambda t y \rangle =
\sum_
{\tiny
\begin{tabular}{c}
\mbox{all different} \\
\mbox{paths from minimal}\\
\mbox{to maximal vertex}\\
\mbox{of the graph}			
\end{tabular}
}
\prod_{
\tiny
\begin{tabular}{c}
\mbox{all edges} \\
\mbox{of the one path}\\
\mbox{of the graph}			
\end{tabular}
}
\left.
 \begin{picture}(1,1)
  \raisebox{2pt}{ \put(0,0){\line(1,3){9}}  \put(0,0){\line(1,-3){9}} }
\end{picture}
\;\;\;
\begin{array}{c}
  \mbox{[$m$]}_n + e_n(\tau_n) \\
  \mbox{[$m$]}_{n-1} + e_{n-1}(\tau_{n-1}) \\
  \vdots \\
   \mbox{[$m$]}_k + e_k(\tau_k)\\
  (m)_{k-1}\\
   \end{array}
 \right |
 \hat F_{k, \tau_n}
\left |
\begin{array}{l@{}}
  [m]_n  \\
  \mbox{[$m$]}_{n-1} \\
 \vdots \\
   \mbox{[$m$]}_k \\
  (m)_{k-1}\\
   \end{array}
 \right. \;\;\;\;\,
 \begin{picture}(1,1)
  \raisebox{2pt}{ \put(0,0){\line(-1,3){9}}  \put(0,0){\line(-1,-3){9}} }
\end{picture}
\end{equation}
  \hrulefill
\end{figure*}
where $\hat F_{k, \tau_n}$  is the fundamental tensor operator, $k=f(j)$ and 
$\tau_n=row(\lambda_{1..j}\setminus~~\lambda_{1..j-1})$.
This formula is based on the analogy to the $n$ slit interference experiment with electrons, where the probability amplitude for the transition of an electron, from a source \emph{s} through a sequence of walls with slits in them to the detector \emph{x}, is given by the formula
$$
\langle x|s\rangle = \sum_{\mbox{{\scriptsize
\begin{tabular}{c}
\mbox{all paths} \\
\mbox{from $s$ to $x$}\\
\end{tabular}
}}} \;\; \prod_
{\scriptsize
\begin{tabular}{c}
\mbox{all parts (edges)} \\
\mbox{of a path}\\
\end{tabular}
}
A_{\scriptsize \mbox{a part of a path}}
$$
where $A_{\scriptsize \mbox{a part of a path}}$ denotes the probability amplitude of transition through a part of a given path.

Fundamental tensor operator in (\ref{wsp}) can be calculated using a technique called \emph{pattern calculus} \cite{louck1,louck4,louck4a}. This is used to determine matrix elements of tensor operators of any unitary groups with the help of symbolic diagrams and appropriate processing rules. It is based on Gelfand-Tsetlin patterns, which are converting many complicated dependencies between the arguments of a vector state into \emph{obvious} geometrical limitations (betweenness conditions). The same limitation applies to tensor operators. 

Louck \cite{louck} has shown that this kind of fundamental tensor operator in (\ref{wsp}) can be calculated using the formula
{
\begin{equation}\label{fso}
\begin{array}{l} %dop
\left.
 \begin{picture}(1,1)
  \raisebox{2pt}{ \put(0,0){\line(1,3){10}}  \put(0,0){\line(1,-3){10}} }
\end{picture}
\;\;\;
\begin{array}{c}
  \mbox{[$m$]}_n + e_n(\tau_n) \\
  \mbox{[$m$]}_{n-1} + e_{n-1}(\tau_{n-1}) \\
  \vdots \\
   \mbox{[$m$]}_k + e_k(\tau_k)\\
  (m)_{k-1}\\
   \end{array}
 \right |
 \hat F_{k, \, \tau_n}
\left |
\begin{array}{l@{}}
  [m]_n  \\
  \mbox{[$m$]}_{n-1} \\
 \vdots \\
   \mbox{[$m$]}_k \\
  (m)_{k-1}\\
   \end{array}
 \right.\;\;\;\;\,
 \begin{picture}(1,1)
  \raisebox{2pt}{ \put(0,0){\line(-1,3){10}}  \put(0,0){\line(-1,-3){10}} }
\end{picture}
\\ \\ \quad \quad \quad \quad \quad \quad \quad \quad \quad \quad \quad \quad \quad = 
\prod_{j=k+1}^{n} \mbox{sgn}(\tau_{j-1} - \tau_j)
\\ \\ %dop
\sqrt{
\left |
\frac
{\mathop{\prod_{i=1}^{j-1}}_{i \neq \tau_{j-1}}
\prod_{i=1, i \neq \tau_{j-1}}^{j-1} (p_{\tau_j,j} - p_{i,j-1})
 \prod_{i=1, i \neq \tau_{j}}^{j} (p_{\tau_{j-1},j-1} - p_{i,j}+1)}
{\prod_{i=1, i \neq \tau_{j}}^{j} (p_{\tau_j,j} - p_{i,j})
 \prod_{i=1, i \neq \tau_{j-1}}^{j-1} (p_{\tau_{j-1},j-1} - p_{i,j-1}+1)}
\right |
}
\\ \\ %dop
\sqrt{
\left |
\frac
{\prod_{i=1}^{k-1} (p_{\tau_k,k} - p_{i,k-1})}
{\prod_{i=1, i \neq \tau_{k}}^{k} (p_{\tau_k,k} - p_{i,k})}
\right |
}
\end{array} %dop
\end{equation}
}
for $k \in \{2,3, \ldots, n-1 \}$. For $k=n$ the first factor
{\scriptsize
$$
\sqrt{
\left |
\frac
{\mathop{\prod_{i=1}^{j-1}}_{i \neq \tau_{j-1}}
\prod_{i=1, i \neq \tau_{j-1}}^{j-1} (p_{\tau_j,j} - p_{i,j-1})
 \prod_{i=1, i \neq \tau_{j}}^{j} (p_{\tau_{j-1},j-1} - p_{i,j}+1)}
{\prod_{i=1, i \neq \tau_{j}}^{j} (p_{\tau_j,j} - p_{i,j})
 \prod_{i=1, i \neq \tau_{j-1}}^{j-1} (p_{\tau_{j-1},j-1} - p_{i,j-1}+1)}
\right |
}$$
}is equal to 1; while for $k=1$ the second factor of the product
{\scriptsize
$$\sqrt{
\left |
\frac
{\prod_{i=1}^{k-1} (p_{\tau_k,k} - p_{i,k-1})}
{\prod_{i=1, i \neq \tau_{k}}^{k} (p_{\tau_k,k} - p_{i,k})}
\right |
}
$$
}is equal to 1.
The partial hook $p_{ij} = m_{ij}+j-i$, $e_i(j)$ is the unit vector of the length $i$ with $1$ on the position $j$,  $[m]_i$ represents $i$-th row of Gelfand-Tsetlin pattern $(m)$, whereas $(m)_i$ denotes rows from 1 to $i$ of pattern $(m)$.

Equation (\ref{fso}) allows to express any matrix element of any fundamental tensor operator in a basis of Gelfand-Tsetlin patterns. 

\section{Example}

Let us calculate the matrix element of the form 
$$
\Big\langle f=(1,3,2,1)      \; \Big | \;       \lambda =(3,1), t~=~{\scriptsize \Yvcentermath1 \young(113,2)}, y~=~{\scriptsize \Yvcentermath1 \young(124,3)}   \Big\rangle,
$$
i.e. element of the Schur-Weyl matrix with the row given by $|1,3,2,1\rangle$ and the column 
\begin{equation*}
| \;       \lambda =(3,1), t~=~{\scriptsize \Yvcentermath1 \young(113,2)}, y~=~{\scriptsize \Yvcentermath1 \young(124,3)}   \Big\rangle
\end{equation*}
for the Heisenberg magnet with $N=4$ nodes and single node spin $s=1$ ($n=3$). 

The Weyl tableau ${\scriptsize \Yvcentermath1 \young(113,2)}$ is in bijection with Gelfand pattern 
$
{\tiny
\left (
\begin{array}{ccccc}
  3 &   & 1 &   & 0  \\
    & 2 &   & 1 &   \\
    &   & 2 &  &  \\
 \end{array}
 \right )}.
$
Appropriate addition, as a one-by-one process, of the single-node states to a zero Gelfand triangle generates a graph of the form

$$
\begin{array}{ccccc}
&
{\tiny
\left (
\begin{array}{ccccc}
  0 &   & 0 &   & 0  \\
    & 0 &   & 0 &   \\
    &   & 0 &  &  \\
 \end{array}
 \right )}

 &&&\\
 \vspace{-3pt}
 &&\\
 \vspace{-3pt}
 &\downarrow {\tiny 1}&\\
 &&\\
 &
{\tiny
\left (
\begin{array}{ccccc}
  1 &   & 0 &   & 0  \\
    & 1 &   & 0 &   \\
    &   & 1 &  &  \\
 \end{array}
 \right )}
 &&&\lambda_1 =(1,0,0) \\
 &&\\
 &\downarrow {\tiny 3}&\\
 &&\\
 &
{\tiny
\left (
\begin{array}{ccccc}
  2 &   & 0 &   & 0  \\
    & 1 &   & 0 &   \\
    &   & 1 &  &  \\
 \end{array}
 \right )}
 &&&\lambda_{12} =(2,0,0) \\
 &&\\
 \swarrow {\tiny 2}&&\searrow {\tiny 2}\\
 &&\\
{\tiny
\left (
\begin{array}{ccccc}
  2 &   & 1 &   & 0  \\
    & 2 &   & 0 &   \\
    &   & 1 &  &  \\
 \end{array}
 \right )}
 &&
 {\tiny
 \left (
\begin{array}{ccccc}
  2 &   & 1 &   & 0  \\
    & 1 &   & 1 &   \\
    &   & 1 &  &  \\
 \end{array}
 \right )}
 &&
 \lambda_{123} =(2,1,0) \\
 &&\\
 \searrow {\tiny 1}&&\swarrow {\tiny 1}\\
 &&\\
 &
{\tiny
\left (
\begin{array}{ccccc}
  3 &   & 1 &   & 0  \\
    & 2 &   & 1 &   \\
    &   & 2 &  &  \\
 \end{array}
 \right )}
 &&&\;\;\;\;\;\; \lambda_{1234} = (3,1,0) \\
\end{array}
$$
~~\\~~\\
adjusted to the set of partitions (taken from the RSK algorithm) presented on the right hand side of this graph. 

Due to having two distinct paths within the graph (from the top to the bottom), we then have the sum of products (according to the equation (\ref{wsp})) of the form

$$
\Big \langle {\scriptsize (1,3,2,1)} \Big | {\scriptsize (3,1)}, {\scriptsize \Yvcentermath1 \young(113,2)}, \; {\scriptsize \Yvcentermath1 \young(124,3)} \Big\rangle =
$$
~~\\
$$
{\scriptsize
\left <
\begin{array}{@{}c@{}c@{}c@{}c@{}c@{}}
  2 &   & 0 &   & 0  \\
    & 1 &   & 0 &   \\
    &   & 1 &  &  \\
 \end{array}
\right |
t_{31}
\left |
\begin{array}{@{}c@{}c@{}c@{}c@{}c@{}}
  1 &   & 0 &   & 0  \\
    & 1 &   & 0 &   \\
    &   & 1 &  &  \\
 \end{array}
\right >
\;
\left <
\begin{array}{@{}c@{}c@{}c@{}c@{}c@{}}
  2 &   & 1 &   & 0  \\
    & 2 &   & 0 &   \\
    &   & 1 &  &  \\
 \end{array}
\right |
t_{22}
\left |
\begin{array}{@{}c@{}c@{}c@{}c@{}c@{}}
  2 &   & 0 &   & 0  \\
    & 1 &   & 0 &   \\
    &   & 1 &  &  \\
 \end{array}
\right >
\;
\left <
\begin{array}{@{}c@{}c@{}c@{}c@{}c@{}}
  3 &   & 1 &   & 0  \\
    & 2 &   & 1 &   \\
    &   & 2 &  &  \\
 \end{array}
\right |
t_{11}
\left |
\begin{array}{@{}c@{}c@{}c@{}c@{}c@{}}
  2 &   & 1 &   & 0  \\
    & 2 &   & 0 &   \\
    &   & 1 &  &  \\
 \end{array}
\right >\; +
}
$$
$$
{\scriptsize
\left <
\begin{array}{@{}c@{}c@{}c@{}c@{}c@{}}
  2 &   & 0 &   & 0  \\
    & 1 &   & 0 &   \\
    &   & 1 &  &  \\
 \end{array}
\right |
t_{31}
\left |
\begin{array}{@{}c@{}c@{}c@{}c@{}c@{}}
  1 &   & 0 &   & 0  \\
    & 1 &   & 0 &   \\
    &   & 1 &  &  \\
 \end{array}
\right >
\;
\left <
\begin{array}{@{}c@{}c@{}c@{}c@{}c@{}}
  2 &   & 1 &   & 0  \\
    & 1 &   & 1 &   \\
    &   & 1 &  &  \\
 \end{array}
\right |
t_{22}
\left |
\begin{array}{@{}c@{}c@{}c@{}c@{}c@{}}
  2 &   & 0 &   & 0  \\
    & 1 &   & 0 &   \\
    &   & 1 &  &  \\
 \end{array}
\right >
\;
\left <
\begin{array}{@{}c@{}c@{}c@{}c@{}c@{}}
  3 &   & 1 &   & 0  \\
    & 2 &   & 1 &   \\
    &   & 2 &  &  \\
 \end{array}
\right |
t_{11}
\left |
\begin{array}{@{}c@{}c@{}c@{}c@{}c@{}}
  2 &   & 1 &   & 0  \\
    & 1 &   & 1 &   \\
    &   & 1 &  &  \\
 \end{array}
\right >\; =
}
$$
$$
{\scriptsize
\left(\frac{\sqrt{2}}{2}\right)\left(-\frac{\sqrt{6}}{6}\right)\left(-\frac{\sqrt{3}}{12}\right) + \left(\frac{\sqrt{2}}{2}\right)\left(\frac{\sqrt{2}}{2}\right)\left(\frac{3}{4}\right)= \; \frac{5}{12},
}
$$
where fundamental tensor operators was calculated in terms of formula (\ref{fso}).

\section{Conclusions and remarks}

We have demonstrated an alternative way of construction of Schur-Weyl transform by  consecutive joining nodes of the spin system according to the symmetry given by the tableaux $(t,y)$. The proposed method is based on graph theory and a technique called pattern calculus. 

The novelty of the algorithm and also the main idea behind it is the fact that while building the quantum state via addition of consecutive nodes we are dealing with a ``combinatorial quantum interference'' of all the possible ways of addition of the new node to an existing system. This observation leads to (\ref{wsp}) which significantly simplifies the procedure of calculating the matrix elements of Schur-Weyl tableau.

All operations being carried out in the proposed algorithm are essentially limited to addition and multiplication since it calculates each amplitude in a polynomial time with parameters $N$ and $n$ in contrary to the standard method \cite{bohr_mottelson} which uses the summation over the symmetric group, and thus grows exponentially with $N$.

%~~\\~~
\section*{Acknowledgements}

We acknowledge the support from the Centre for Innovation and Transfer of Natural Sciences and Engineering Knowledge at the University of Rzesz\'ow.

\end{document}